\documentclass[a4paper]{jpconf}
\pdfoutput=1
\usepackage{graphicx,xcolor}
\usepackage{rotating,booktabs,amsmath,amssymb, multicol,url}
\usepackage[numbers,sort&compress]{natbib}
\usepackage{natbibspacing}
\usepackage{ulem}
\usepackage{multicol}
\begin{document}
\title{Targeting the Conformal Window:\newline Scalars on the Lattice}

\author{Evan Weinberg}
\address{ Department of Physics, Boston University, Boston, MA, USA}
\ead{weinbe2@bu.edu}

\author{Rich Brower}
\address{Department of Physics, Boston University, Boston, MA, USA}
\ead{brower@bu.edu}

\author{Anna Hasenfratz}
\address{Department of Physics, University of Colorado Boulder, Boulder, CO, USA}
\ead{Anna.Hasenfratz@colorado.edu}

\author{Claudio Rebbi}
\address{Department of Physics, Boston University, Boston, MA, USA}
\ead{rebbi@bu.edu}

\author{Oliver Witzel\footnote{Present address: School of Physics \& Astronomy, The University of Edinburgh, EH9 3FD, UK}}
\address{ Center for Computational Science, Boston University, Boston, MA, USA}
\ead{owitzel@bu.edu}

\begin{abstract}
The light Higgs boson of the Standard Model could arise as the consequence of the weakly broken
conformal symmetry in a strongly interacting gauge theory. Here we
present a novel idea to study the transition from conformal to confining behavior using an $SU(3)$ gauge
theory with four light and eight heavy flavors. This system interpolates between the 12-flavor conformal
and the 4 flavor chirally broken theory as the mass of the heavy flavors are varied. We show first results
on our determination of the iso-singlet $0^{++}$ state.
\end{abstract}

\section{Introduction}
\label{Intro}
In 2012 the Higgs boson was discovered at the LHC \cite{Aad:2012tfa,Chatrchyan:2012ufa} adding the final missing piece to the electroweak sector of the Standard Model. In the coming years the experiments will  improve their precision and resolve further properties of the Higgs particle. It is of particular interest is to learn more about the origin of the Higgs boson and the nature of electroweak symmetry breaking. The  fundamental, self-interacting scalar as described by the Standard Model poses a theoretical challenge because it is not ultraviolet complete. One viable scenario is  to consider the Higgs boson as a composite particle with its own underlying strongly interacting but chirally broken gauge theory~\cite{Yamawaki:1985zg,Appelquist:1991nm}. In such composite Higgs models the Higgs boson arises as  a fermionic $0^{++}$ iso-singlet bound state. 

QCD-like systems have a plethora of low-energy bound states and we expect to find a similar spectrum in scaled-up QCD models.   Viable composite Higgs models have to be very different, predicting that the  mass of any non-Goldstone states  is much heavier than the iso-singlet scalar. In addition to an unusual spectrum, composite Higgs models  have to satisfy other constraints, like predicting the experimentally observed Higgs decay width or the S-parameter correctly. These constraints could be met by a strongly interacting theory with a weakly broken conformal symmetry  --- a conjecture which can be numerically tested using lattice gauge theory. 

Recent lattice results look very promising:  there is increasing evidence for an infrared fixed point (IRFP) for 12 fundamental flavors \cite{Appelquist:2011dp,Hasenfratz:2011xn,Aoki:2012eq,DeGrand:2011cu,Cheng:2013xha,Cheng:2014jba,Lombardo:2014pda,Aoki:2013zsa} and two groups have reported on a low mass scalar in the SU(3) gauge  theory: one for  8 fundamental flavors \cite{Aoki:2014oha}, the other for   2 sextet flavors \cite{Fodor:2014pqa}.  These investigations are  limited by an integer flavor number. {\it A priori}\/ it is not obvious that a theory with an integer fermion number will be close enough to the conformal window where the IRFP appears. Also there are many more gauge groups or fermion representations to be considered. To facilitate such a search and overcome the limitation of integer flavor numbers, we proposed to simulate a theory with $N_\ell=4$ light and $N_h=8$ heavy flavors \cite{Brower:2014dfa}. Keeping the four light flavors near the chiral limit, i.e.~$m_\ell \approx 0$, we use the eight heavy flavors with mass $m_h \ge m_\ell$ to interpolate between the $N_\ell=4$ chirally broken theory and the $N_h+N_\ell=12$ flavor mass-deformed conformal theory. Phenomenlogically it may be more interesting to simulate a model with, e.g., 2+1 flavors of sextet fermions since this has exactly three massless Goldstone bosons as needed by electroweak symmetry breaking. Our choice of 4+8 flavors is  motivated by our use of the staggered formulation to discretize the fermions which has a multiple of four as natural choice. Furthermore, the above mentioned results indicate that the limiting cases of four and twelve flavors have the desired properties, implying  4+8 flavors should certainly interpolate in the relevant parameter space, although the anomalous dimension of the $SU(3)$ 12-flavor system may be too small $(\gamma_m^\star \approx 0.24)$ \cite{Cheng:2013eu,Cheng:2013xha,Lombardo:2014pda} to satisfy phenomenological walking constraints.

In our first numerical studies presented at Lattice 2014 \cite{Brower:2014ita}, we demonstrated that the renormalized running coupling in our setup  becomes a ``walking'' coupling in an intermediate energy range.  Having a walking regime is necessary to satisfy known phenomenological constraints. We also found that the energy range of walking increases as we lower the mass of the eight heavy flavors, i.e., as we approach the IRFP in the 12-flavor theory, confirming our expectations. Next we study the impact of a walking gauge coupling on the meson spectrum. While the computation of the (connected) meson spectrum of states like the pion, rho, or $a_0$ is standard, the computation of the $0^{++}$ sigma particle brings additional challenges because it requires the evaluation of disconnected contributions. Disconnected contributions arise from quark-anti-quark pairs coupled only via a gluon to the state of interest.  The evaluation of those is numerically much more demanding.  In this paper we focus on  the computation of the  mass of the $0^{++}$ meson  and show preliminary results for the meson spectrum of our 4+8 flavor theory. 

In the following section we introduce the setup of our simulations and provide details on the generated ensembles of gauge field configurations. Next we discuss in Section \ref{SecMesSpectrum} the details and results of our computation of the connected and disconnected spectrum, before we finally conclude.

\section{Numerical setup}
\label{SecNumSetup}
\begin{figure}[tb]
{\begin{picture}(149, 57)
   \put(0,1){\includegraphics[width=0.49\textwidth]{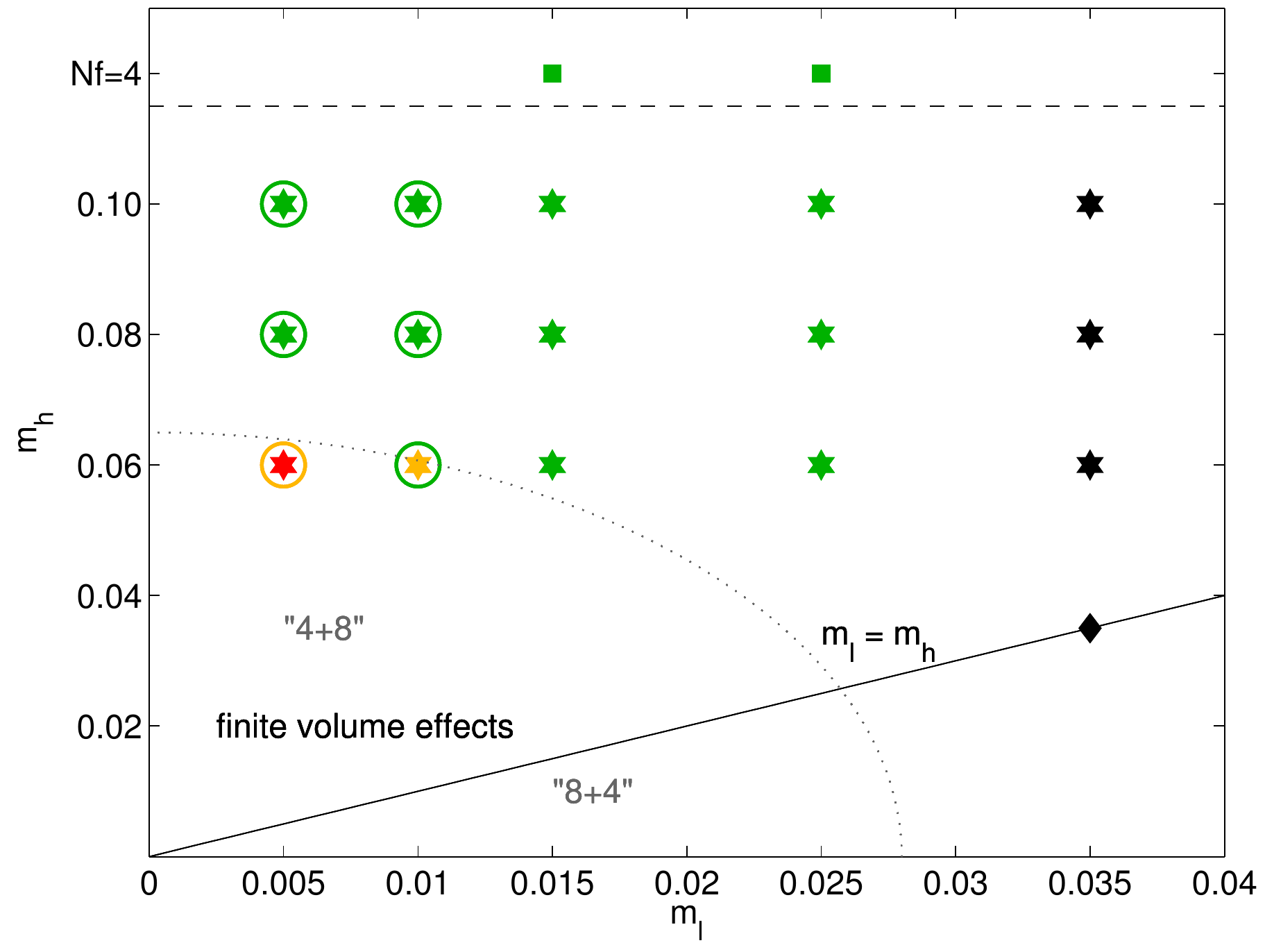}}
   \put(76,0){ \includegraphics[width=0.48\textwidth]{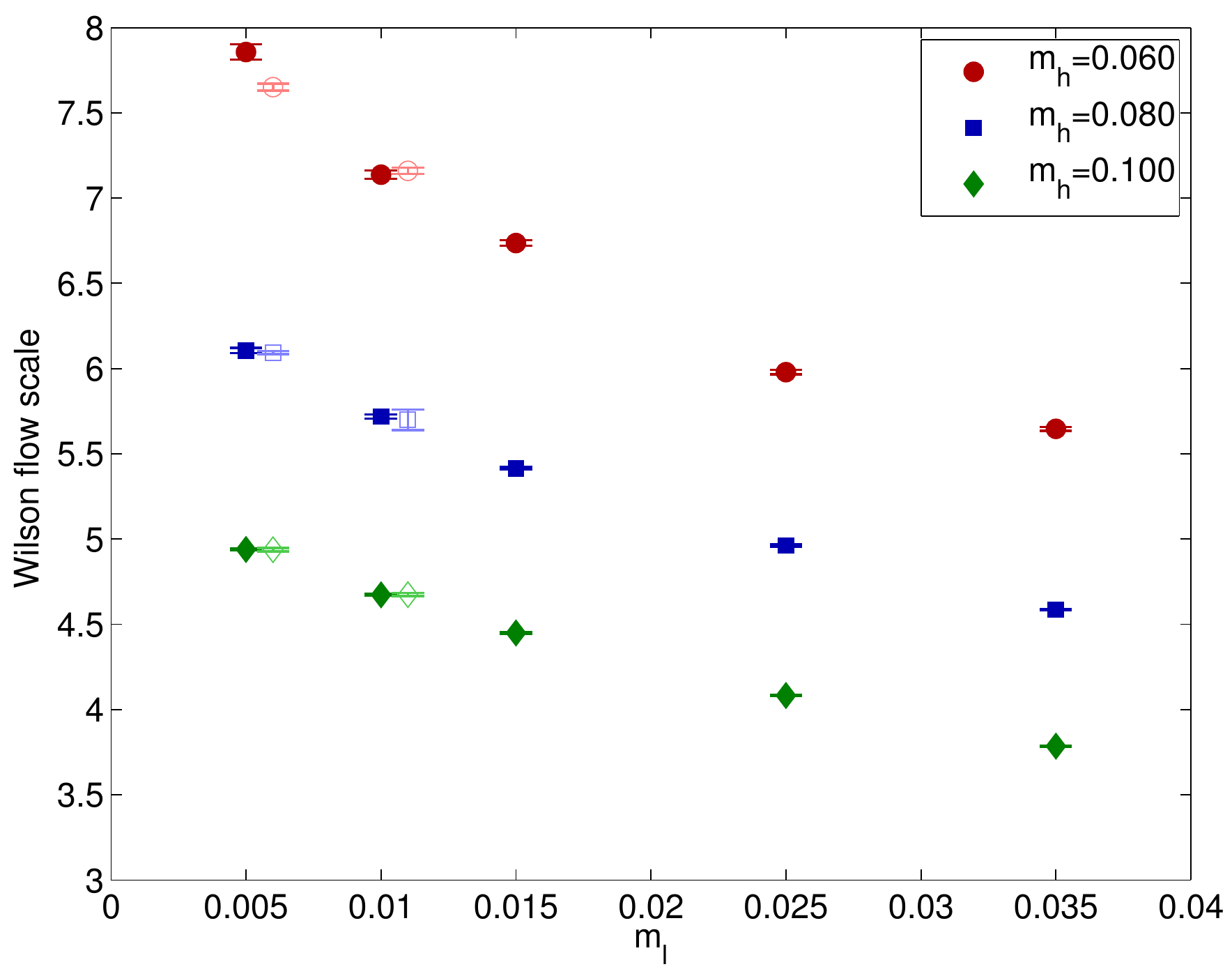}}
   \put(77,39){\rotatebox{90}{\tiny$\mathsf{\sqrt{8\tilde t_0}}$}}
\end{picture}}
    \caption{Left panel: light ($m_\ell$) and heavy ($m_h$) mass values for the simulations carried out on $24^3\times 48$ lattices (filled symbols) and $32^3\times 64$ lattice (open circles). The colors are meant to caution about finite size effects, likely negligible for green, but of increasing importance as the color turns to orange and red. The black data points are likely to heavy and troubled by cut-off effects.
Right panel:  The gradient flow scale $\sqrt{8\tilde{t_0}}$ for our 21 ensembles vs.~$m_\ell$. Filled symbols show values determined on $24^3\times 48$ lattices, open symbols (shown with a small horizontal offset) refer to the values measured on $32^3\times 64$ lattices. The strong dependence on both the heavy and light fermion masses is most likely the effect of the IRFP of the 12 flavor system.}
    \label{f2}
\end{figure}

We use a lattice action based on nHYP \cite{Hasenfratz:2007rf} smeared staggered fermions and the plaquette gauge action containing fundamental and adjoint terms \cite{Cheng:2013xha,Cheng:2013bca}. Gauge field ensembles are generated using the hybrid Monte Carlo (HMC) update algorithm \cite{Duane:1987de} as implemented in the FUEL software package \cite{FUEL}. The combined choice of the gauge and fermion actions  has been extensively tested in projects using various number of fermion flavors (see e.g.~\cite{Cheng:2013xha,Cheng:2013bca}) and proven to provide numerically stable simulations for our choice of parameters. Moreover, taste breaking effects of staggered fermions are largely suppressed when nHYP smeared. So far we generated ensembles with a range of different light masses $m_\ell$ and heavy masses $m_h$, analyzing two different volumes $24^3\times 48$ and $32^3 \times 64$ at one value of the gauge coupling ($\beta=4.0$, $\beta_a = -\beta/4$ \cite{Hasenbusch:2004yq}). In this paper we focus on the 4+8 flavor ensembles summarized in the left panel of Fig.~\ref{f2}, a total of 21 different ensembles at five different values of the light quark mass $m_\ell$ and three different values of the heavy quark mass $m_h$. Work is still in progress to extend some of the existing runs and add new ensembles at different light or heavy quark masses. The color coding in the above mentioned plot highlights in green, orange and red those ensembles which are most useful for our determinations of the $0^{++}$ scalar with the colors indicating finite size effects (negligible for green and severe for red). As expected we observe finite size effects becoming significant when reducing the masses. Simulations with $m_\ell=0.035$ turned out to have a too large light mass to reliably extract  the $0^{++}$ and are not  considered in our analysis.

The right plot in Fig.~\ref{f2} shows the lattice scale of our ensembles and how it depends on the light and heavy quark mass. We determine the lattice scale using the concept of gradient or Wilson flow \cite{Narayanan:2006rf,Luscher:2009eq,Luscher:2010iy} which defines the quantity $t_0$. In order to reduce $\mathcal{O}(a^2)$ lattice artifacts we consider the t-shift improved gradient flow by shifting the Wilson flow parameter $t$ by a small constant $\tau_0$ \cite{Cheng:2014jba}. Empirically we find that $\tau_0=0.1$ is close to optimal and removes nearly all cut-off effects  of $t_0$ \cite{Brower:2014ita}. We denote our $t$-shifted scale by $\tilde t_0$ and use the quantity  $\sqrt{8 \tilde t_0}$ to compare the lattice scales of our different ensembles as shown  in the right plot of Fig.~\ref{f2}. Reducing cut-off effects is important since currently we only have data at one $\beta$ value which does not allow us to take a proper continuum limit. As is shown  in the plot, our scale has a strong, non-linear dependence on $m_\ell$ and changes significantly with $m_h$. Simulating at lower values of $m_\ell$, as needed to approach  the chiral limit, will therefore force us to simulate on larger volumes. As a rule of thumb, values of  $\sqrt{8 \tilde t_0} \lesssim L/5$ are typically considered to be safe from  finite size effects as verified by  comparing the full ($24^3\times 48$) and open ($32^3\times 64$) symbols. We observe a significant difference between the two volumes only for the simulation with the lightest input masses, $m_\ell = 0.005,\, m_h=0.060$.

\section{Calculating the Light flavor spectrum}
\label{SecMesSpectrum}

As mentioned above, the iso-singlet $0^{++}$ meson receives special attention in models beyond the Standard Model as it could be a Higgs candidate. Its mass  is considerably more difficult to measure numerically than other, non-singlet mesons because both quark-line connected and disconnected diagrams contribute to the correlator
\begin{align}
C_{0^{++}}(t) &\equiv \frac{N_\ell}{4}C_{\rm{disc} }(t) - C_{\rm{conn}}(t).
\label{eq:combined}
\end{align}
 The disconnected contribution is given by
\begin{align}
C_{\rm{disc}}(t)=
 \sum_{t_0} \left(\langle \bar{\psi} \psi \rangle(t_0) - \langle\langle\bar{\psi}\psi\rangle\rangle_{\rm{e}}\right)\left(\langle \bar{\psi} \psi \rangle(t_0+t) - \langle\langle\bar{\psi}\psi\rangle\rangle_{\rm{e}}\right)
\label{eq:disc}
\end{align}
where  $\langle\langle\bar{\psi}\psi\rangle\rangle_{\rm{e}}$ denotes the ensemble average of the fermion condensate. Accurately measuring the ensemble average of the condensate is essential for these calculations as large errors in  $\langle\langle\bar{\psi}\psi\rangle\rangle_{\rm{e}}$ can destroy the correlator of Eq.~(\ref{eq:disc}) at large $t$.

Further complicating matters, only a stochastic estimate of the matrix element $ \langle \bar{\psi} \psi \rangle(t)$ is feasible. Care must be and is taken to ensure no bias is introduced by incorrectly combining stochastically measured quantities. We compute it on each considered configuration using $N_r$  $U(1)$ noise sources $\eta_i$ with the property 
\begin{equation}
\lim_{N_r \rightarrow \infty} \frac{1}{N_{r}} \sum_{i} \eta^\dagger_{i}(\vec{x},t) \eta_{i}(\vec{y},t') = \delta_{\vec{x},\vec{y}} \delta_{t,t'}.
\end{equation}
 Inverting the staggered Dirac operator $D((\vec{x},t), (\vec{y},t'))$  on the noise source $\eta_i$ gives
 \begin{equation}
  \phi_{i}(\vec{x},t) = D((\vec{x},t),(\vec{y},t'))^{-1} \eta_{i}(\vec{y},t'),
 \end{equation}
  and we can compute the desired matrix element  as
\begin{align}
\langle \bar{\psi} \psi \rangle(t) &= \lim_{N_r\rightarrow \infty} \frac{1}{N_r} \sum_i \sum_{\vec{x}}\eta_i^\dagger(\vec{x},t) \phi_i(\vec{x},t). 
\label{eqn:pbp}
\end{align}
 In order to enhance the signal, we use $N_r=6$ noise sources spreading over the full 4-d volume of the lattice diluted in time, color, as well as even/odd in space~\cite{Foley:2005ac}. This gives a reasonable balance between numerical costs and the signal-to-noise ratio. Our dilution pattern is consistent with the similar calculation performed in Ref.~\cite{Fodor:2014pqa}.  Moreover, we use an improved operator for the chiral condensate, unique to na\"ive and staggered fermions, which further reduces variance in our measurement~\cite{Kilcup:1986dg}.

The above parameter choices require 1728 inversions of a single color fermion matrix for each $24^3\times 48$ configuration, and 2304 inversions for each $32^3\times 64$ configuration. Using our full ensembles we are able to determine the  disconnected correlators to a few percent precision this way and extract the mass of the corresponding  $0^{++}$ states.
  For comparison,  for a $24^3\times 48$ ($32^3\times 64$) lattice volume, it suffices to perform 18 (24) inversions of a single color fermion matrix to evaluate the quantities of the connected spectrum, e.g., the pion and rho meson, using wall sources to determine their mass with a precision to better than a percent on our ensembles. 
This illustrates the cost and difficulties  of disconnected correlator  calculations.

\begin{figure}[tb]
{\begin{picture}(149, 57)
\put(0,0){\includegraphics[width=0.5\textwidth]{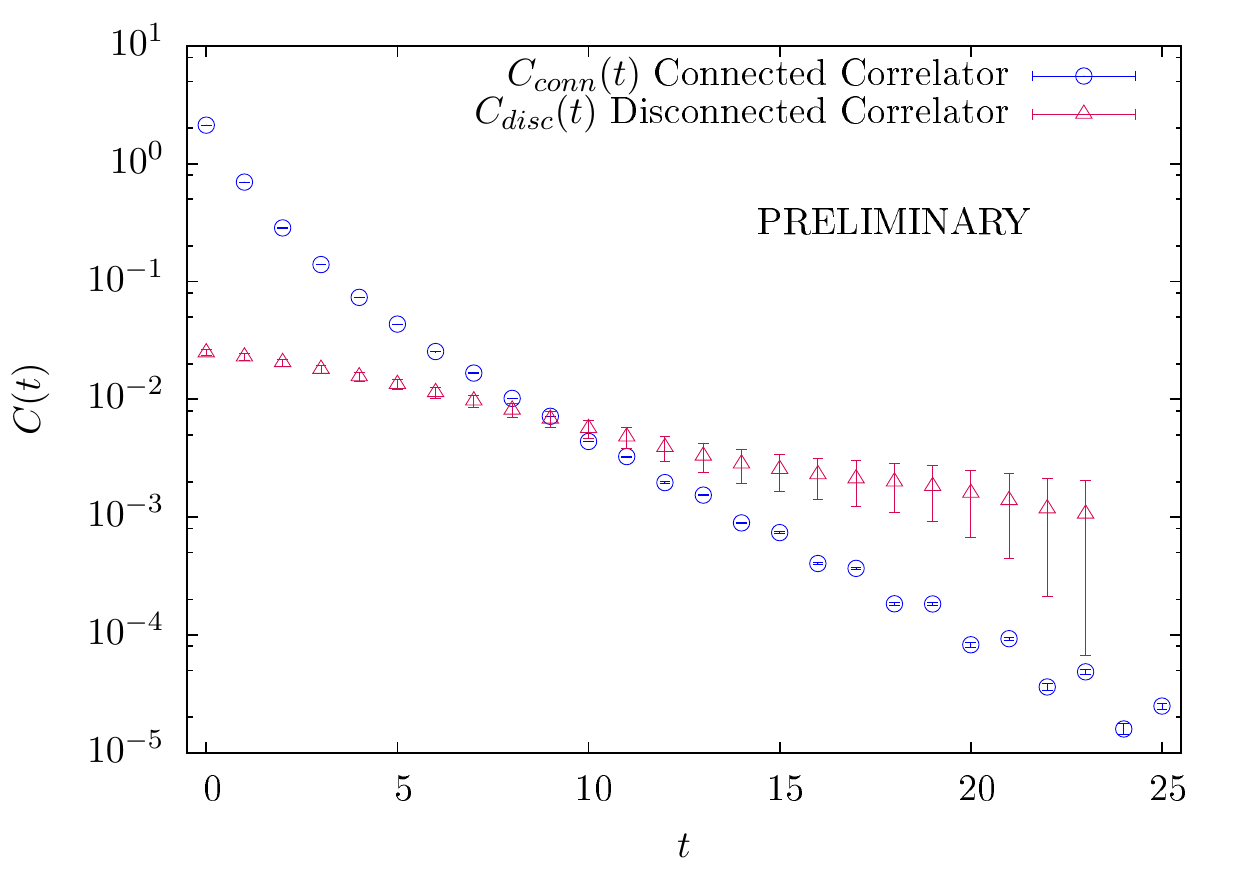}}
\put(76,0){\includegraphics[width=0.5\textwidth]{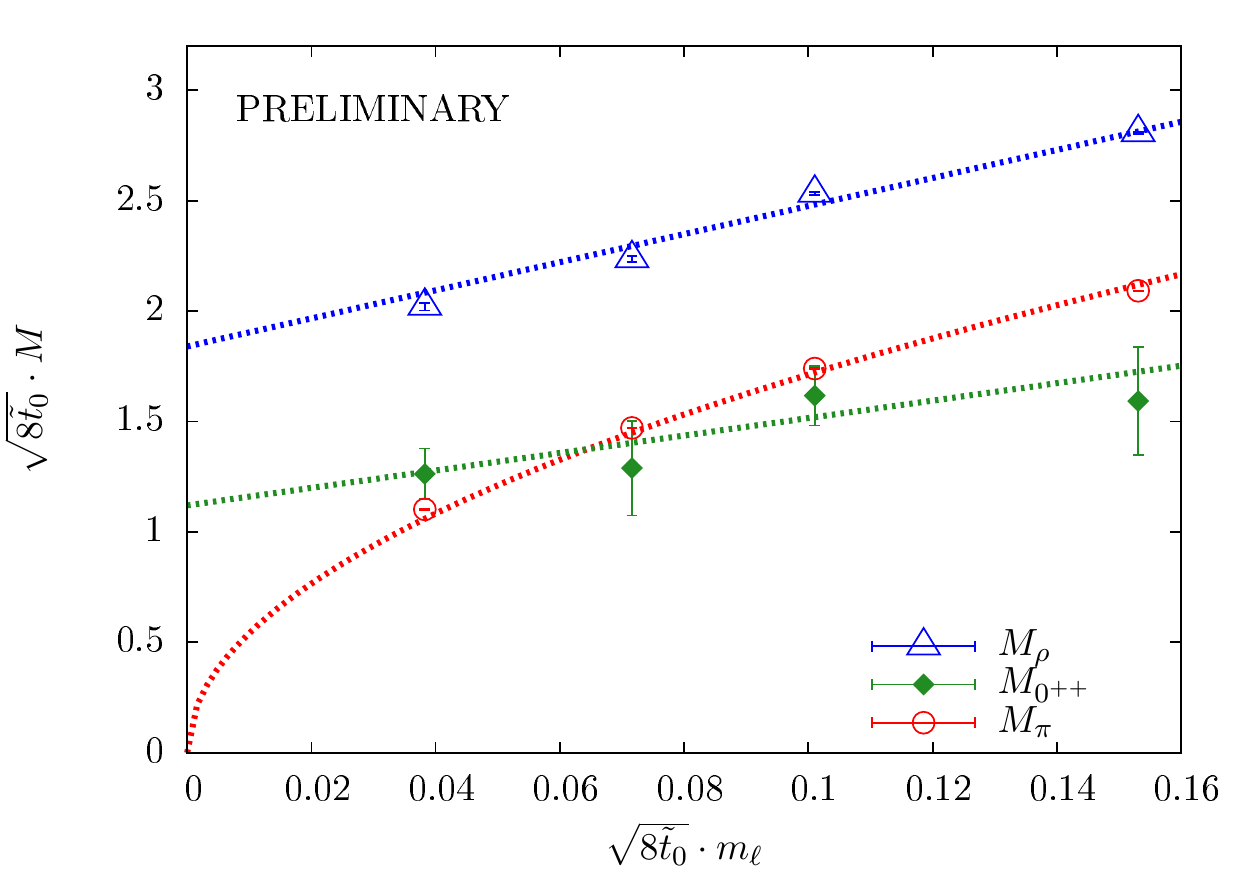}}
\end{picture}}
\caption{Left panel: Comparison of $-C_{\rm{conn}}(t)$ and $C_{\rm{disc}}(t)$ for the $32^3\times 64$ $m_\ell = 0.010$. Right panel: The pion, rho and $0^{++}$ scalar light flavor spectrum for the $m_h=0.060$ ensembles. }
\label{fig:correlator}
\end{figure}

As an example we show the resulting connected and disconnected scalar correlators on the $32^3\times64$, $m_\ell=0.010$, $m_h=0.060$ ensemble in the  left panel of Fig.~\ref{fig:correlator}.  Due to the much higher costs we have presently measured the disconnected correlator on only 89 configurations, with  further measurements in progress. 

General features of the $C_{0^{++}}$ correlators are nonetheless present: For small $t$ the connected correlator dominates the combination  and the effect of the disconnected contribution is insignificant until $t\gtrsim 8$. There the signal of $C_{\rm{disc}}$ is very noisy and we are not able to fit the mass reliably. As an alternative Refs.~\cite{Aoki:2013zsa,Aoki:2014oha,Fodor:2014pqa} advocate to consider the disconnected correlator only. As long as the $0^{++}$ state is lighter than the non-singlet $a_0$ state,  the iso-singlet scalar mass can be extracted directly from $C_{\rm{disc}}$. In practice this works surprisingly well and one heuristic explanation may be that excited states in the connected and the $0^{++}$ correlators cancel. Using $C_{\rm{disc}}$ only we obtain values in agreement with $C_{0^{++}}$ but with smaller statistical uncertainty.

We close this section by showing in the right panel of Fig.~\ref{fig:correlator} preliminary results for the light flavor pion, the rho, and the $0^{++}$ scalar,  obtained on our 4+8 ensembles with $m_h=0.060$ using $32^3\times 64$ volumes for $m_\ell = 0.005$ and $0.010$ and $24^3\times 48$ for $m_\ell=0.015$ and $0.025$. The data shown are rescaled by $\sqrt{8 \tilde t_0}$ to account for differences in the lattice scale. We observe the $0^{++}$ mass that is lighter that the pion at large $m_\ell$ to cross the pion mass and become heavier toward the  chiral limit. This result is of great phenomenological interest and has thus far not been observed in other 
studies of conformal or near-conformal systems \cite{Aoki:2013zsa,Aoki:2014oha,Fodor:2014pqa}. We like however to caution on the preliminary nature of our current data; we are still performing additional measurements and analysis of  data for the $0^{++}$ scalar on other ensembles which are needed to confirm our findings mentioned above.

\section{Conclusions and outlook}
Exploring the phenomena of gauge-fermion systems near the conformal window is interesting in its own and may in addition reveal new ``physics'' which may find its application in beyond the Standard Model theories describing e.g.~the Higgs boson as composite particle. We study a novel model of four light and eight heavy flavors which allows us to tune arbitrarily close to the conformal fixed point of the 12-flavor system while still being chirally broken in the infrared limit. Of special interest is determining the mass of  the iso-singlet scalar which could be a Higgs candidate. This computation is in particular challenging because its measurement contains both a connected and a disconnected contribution.  Using stochastic estimates for the disconnected contribution, our preliminary results reveal the very interesting case of an iso-singlet scalar which is lighter than the pion at large fermion mass, but  heavier for  smaller masses. Our current findings will have to be confirmed by further simulations.

\section*{Acknowledgments}
\label{acknowledgements}
Computations for this work were carried out in part on facilities of
the USQCD Collaboration, which are funded by the Office of Science of the U.S. Department of Energy, on computers at the MGHPCC, in part funded by the National Science Foundation, and on computers allocated under the NSF Xsede program to the project TG-PHY120002. \\
We thank Boston University, Fermilab, the NSF and the U.S. DOE for providing the facilities essential for the completion of this work. R.C.B., C.R. and E.W. were supported by DOE grant DE-SC0010025. In addition, R.C.B., C.R. and O.W. acknowledge the support of NSF grant OCI-0749300. A.H. acknowledges support by the DOE grant DE-SC0010005.

{\small
\bibliography{../General/BSM}
\bibliographystyle{iopart-num}
}

\end{document}